\documentclass[aps,pra,showpacs,twocolumn, superscriptaddress, longbibliography]{revtex4-2}
\usepackage{lineno}
\usepackage{xcolor}
\usepackage[normalem]{ulem} 
\usepackage{graphicx}
\usepackage{dcolumn}
\usepackage{bm}
\usepackage[colorlinks]{hyperref}
\usepackage{amsmath,braket}
\usepackage{mathtools}

\newcommand{\rev}[1]{#1} 

\makeatletter
\renewcommand\@makefnmark{\hbox{\@textsuperscript{\normalfont\@thefnmark}}}
\makeatother

\begin{document}
\title{A hybrid quantum interferometer bridging Hong-Ou-Mandel and N00N-state interference}

\author{Zi-Qi Zeng}
\affiliation{Key Laboratory of Low-Dimensional Quantum Structures and Quantum Control of Ministry of Education, Department of Physics and Synergetic Innovation Center for Quantum Effects and Applications, Hunan Normal University, Changsha 410081, China}
\affiliation{Hubei Key Laboratory of Optical Information and  Pattern Recognition,  Wuhan Institute of Technology, Wuhan 430205, China\\}

\author{Samuel Corona-Aquino}
\affiliation{Hubei Key Laboratory of Optical Information and  Pattern Recognition,  Wuhan Institute of Technology, Wuhan 430205, China\\}
\affiliation{Instituto de Ciencias Nucleares, Universidad Nacional Autónoma de México, Apartado Postal 70-543, 04510 CDMX, Mexico\\}

\author{Tao Xie}
\affiliation{Key Laboratory of Low-Dimensional Quantum Structures and Quantum Control of Ministry of Education, Department of Physics and Synergetic Innovation Center for Quantum Effects and Applications, Hunan Normal University, Changsha 410081, China}

\author{Bin Yin}
\affiliation{Key Laboratory of Low-Dimensional Quantum Structures and Quantum Control of Ministry of Education, Department of Physics and Synergetic Innovation Center for Quantum Effects and Applications, Hunan Normal University, Changsha 410081, China}

\author{Yan Guo}
\affiliation{Hubei Key Laboratory of Optical Information and  Pattern Recognition,  Wuhan Institute of Technology, Wuhan 430205, China\\}

\author{Yukun Song}
\affiliation{State Key Laboratory of Crystal Materials, Shandong University, Jinan 250100, China}

\author{Yun Meng}
\affiliation{School of Precision Instrument and Optoelectronic Engineering, Tianjin University, Tianjin 300072, China}
\affiliation{Key Laboratory of Optoelectronic Information Science and Technology, Ministry of Education, Tianjin 300072, China}

\author{Kai Zou}
\affiliation{School of Precision Instrument and Optoelectronic Engineering, Tianjin University, Tianjin 300072, China}
\affiliation{Key Laboratory of Optoelectronic Information Science and Technology, Ministry of Education, Tianjin 300072, China}

\author{Song Li}
\affiliation{School of Precision Instrument and Optoelectronic Engineering, Tianjin University, Tianjin 300072, China}
\affiliation{Key Laboratory of Optoelectronic Information Science and Technology, Ministry of Education, Tianjin 300072, China}
\affiliation{College of Information Science and Electronic Engineering, Zhejiang University, Hangzhou 310027, China}

\author{Xiaolong Hu}
\affiliation{College of Information Science and Electronic Engineering, Zhejiang University, Hangzhou 310027, China}
\affiliation{State Key Laboratory of Reliability Technology for Electronic Components, Zhejiang University, Hangzhou 310027, China}

\author{Alfred B. U’Ren}
\affiliation{Instituto de Ciencias Nucleares, Universidad Nacional Autónoma de México, Apartado Postal 70-543, 04510 CDMX, Mexico\\}

\author{Roberto de J. León-Montiel}
\affiliation{Instituto de Ciencias Nucleares, Universidad Nacional Autónoma de México, Apartado Postal 70-543, 04510 CDMX, Mexico\\}

\author{Chenglong You}
\affiliation{Institute of Fundamental and Frontier Sciences,University of Electronic Science and Technology of China, Chengdu, Sichuan, 611731, China\\}
\affiliation{Key Laboratory of Quantum Physics and Photonic Quantum Information, Ministry of Education, University of Electronic Science and Technology of China, Chengdu 611731, China\\}

\author{Baihong Li}
\email{Corresponding author:li-baihong@163.com}
\affiliation{School of Physics and Information Science, Shaanxi University of Science and Technology, Xi’an 710021, China\\}

\author{Dongzhou Wang}
\email{Corresponding author:wangdongzhou@jiqt.org }
\affiliation{ Jinan Institute of Quantum Technology and Jinan Branch of Hefei National Laboratory, Jinan 250101, China}

\author{Rui-Bo Jin}
\email{Corresponding author:jrb@hunnu.edu.cn}

\affiliation{Key Laboratory of Low-Dimensional Quantum Structures and Quantum Control of Ministry of Education, Department of Physics and Synergetic Innovation Center for Quantum Effects and Applications, Hunan Normal University, Changsha 410081, China}
\affiliation{Hubei Key Laboratory of Optical Information and  Pattern Recognition,  Wuhan Institute of Technology, Wuhan 430205, China\\}

\date{\today}
\begin{abstract}
Hong-Ou-Mandel (HOM) interference offers robustness in high-noise environments due to its phase-insensitive nature, while N00N-state interference provides enhanced phase sensitivity for precision measurements in low-noise settings. However, their realizations have typically required separate experimental platforms. Here, we report a single hybrid quantum interferometer that can dynamically transition between HOM and N00N-state interference by adjusting two time delays, $\tau_1$ and $\tau_2$. This tunability enables operation in phase-insensitive, phase-sensitive, or hybrid modes within a unified setup, as confirmed by theoretical analysis and experimental demonstration. Our system offers a versatile platform with potential for quantum sensing applications, paving the way for integrated quantum technologies capable of adapting to diverse noise conditions.
\end{abstract}

\maketitle

\section{Introduction}
\begin{figure*}[!thp]
\centering
\includegraphics[width= 0.9\textwidth]{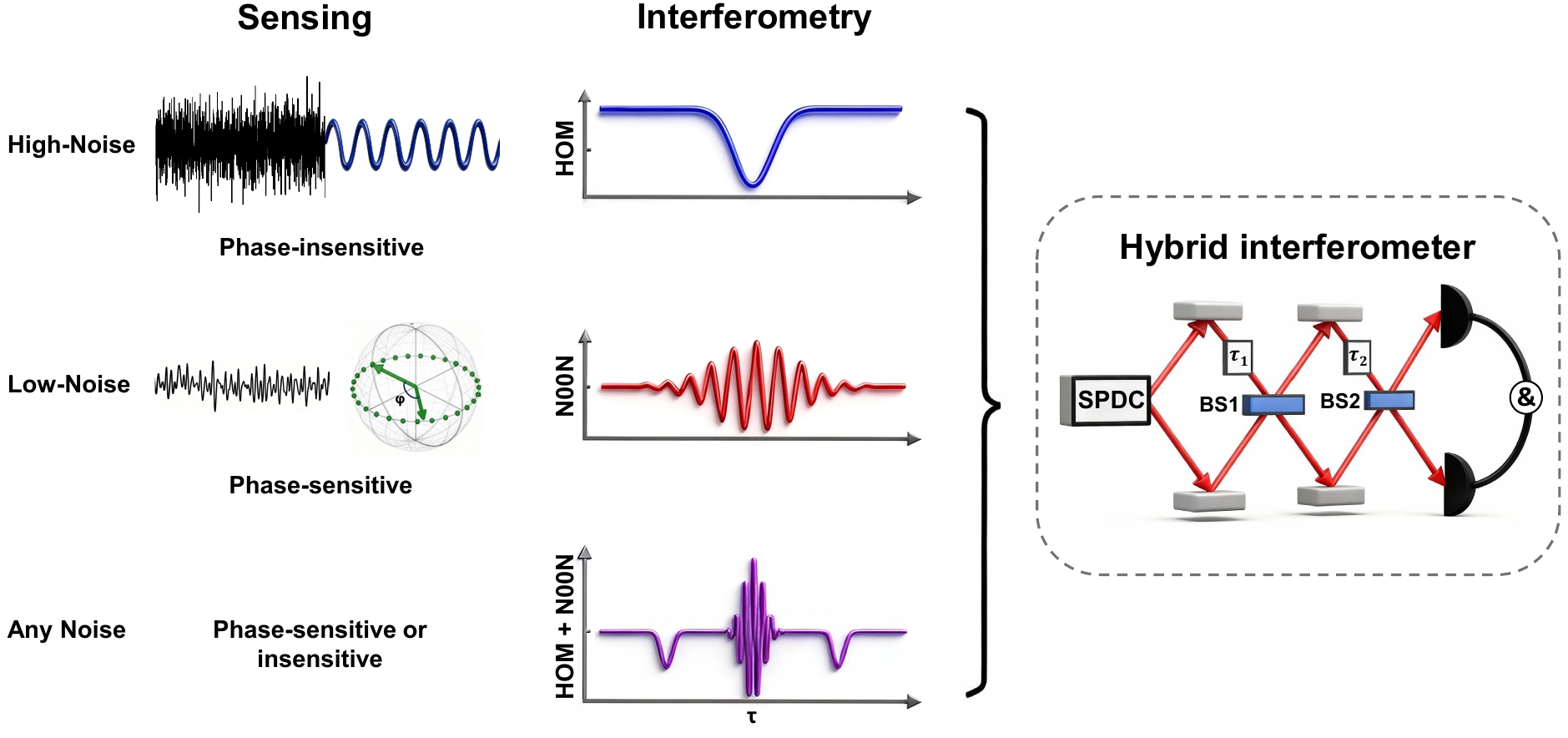}
\caption{Conceptual illustration depicting quantum sensing requirements under different noise conditions (left), their corresponding interferometric approaches (middle), and the hybrid interferometer (right), can simultaneously meet the full demands.}
\label{concept}
\end{figure*}

Quantum sensing plays a significant role in modern science and technology \cite{Degen2017RMP,Pirandola2018NP,Crawford2021AQT,Jiao2025FR}. It exploits the unique properties of quantum systems—such as superposition, entanglement, squeezing, and quantum interference—to achieve measurements whose sensitivity and precision surpass classical limits \cite{Wang2024APR,Zhang2025LPR}. Owing to these remarkable capabilities, quantum sensing has found widespread applications in diverse fields including high-precision magnetometry \cite{Xie2020PhysRevApplied,Lei2024AQT,Gerhard2024PRXQ}, biological microscopy \cite{Taylor2013NP,Casacio2021Nature,Aslam2023NRP}, gravitational-wave detection \cite{Danilishin2019LRR,Wu2023PRD,Tobar2024NC}, and other advanced metrological tasks \cite{Dowling2015JLT,Fiderer2018NC,DeMille2024NP,Jin2024SCPMA}. 

Despite its vast potential, quantum sensing faces the challenge of environmental adaptability, particularly to varying noise levels \cite{Layden2018npj,Lawrie2019ACS,Zhang2019PRL}, as illustrated in Fig.\,\ref{concept}. In practical scenarios, quantum sensors may need to operate in either high- or low-noise environments, each imposing distinct demands on the sensing methodology \cite{Peter2011PRL}. High-noise environments require tools that are highly robust against noise and disturbances \cite{Zhang2015PRL,Salhov2014PRL}, whereas low-noise environments enable the pursuit of ultra-precise measurements, such as super-resolution and super-sensitivity \cite{Boss2017Sicence,Zopes2017PRL,Gefen2019NC}. Quantum interferometers are among the most widely used tools for quantum sensing, and they can be configured to suit these different regimes \cite{Giovannetti2022PRL,Jin2024PQE,Scott2021PRA}. For instance, in high-noise environments, Hong-Ou-Mandel (HOM) interference is often utilized due to its strong noise-resistance capabilities \cite{Lyons2018SA,Jin2022APL,Faurby2024PRL,Jing2025APLPho, Li2025PRL}. Conversely, in low-noise environments, N00N-state interference is preferred, as it enables phase measurements at the Heisenberg limit, achieving much higher precision than the standard quantum limit \cite{Zhou2017PRAppl,Hong2021NC,Hiekkam2021PRL,Hong2022LPR,Kim2025PRL}. However, conventional interferometers are typically confined to operating in a single mode, either in the HOM or the N00N regime \cite{Hong1987,Edamatsu2002PRL}. This limitation restricts their ability to meet diverse sensing needs that require dynamic switching or combination of both interference types within a single device.

To address these limitations and fulfill the diverse requirements of practical quantum sensing, we introduce a novel quantum interferometer capable of achieving both HOM and N00N interferences, as well as their hybridization within a single setup. By carefully adjusting the time delays, specifically $\tau_1$
and $\tau_2$, the interferometer can be continuously adjusted to exhibit phase-insensitive HOM interference, phase-sensitive N00N-state interference, or a hybrid regime that simultaneously manifests both. This tunability allows the device to operate optimally in either high- or low-noise environments, as well as in hybrid scenarios demanding both robustness and super-resolution. The controllable transition and combination of these two distinct quantum interference paradigms introduce a new degree of freedom in photonic quantum engineering. Consequently, our interferometer meets the broad demands of quantum sensing across varying noise conditions, paving the way for more versatile and powerful quantum technologies.

\begin{figure*}[!thp]
\centering
\includegraphics[width= 0.85\textwidth]{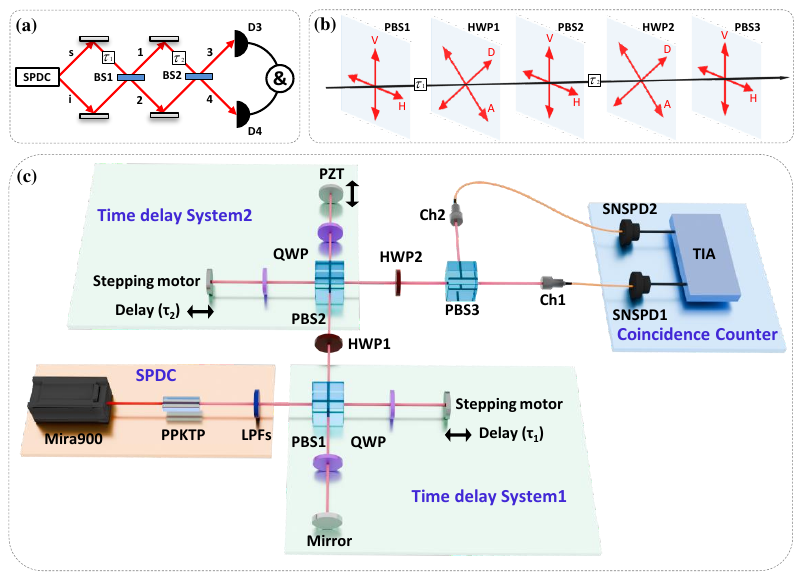}
\caption{(a) The spatial model of the hybrid interferometer. (b) The polarization model for the control of the hybrid interferometer. (c) Experimental setup. LPFs: long-pass filters and dichroic mirror; PBS: polarization beam splitter; QWP: quarter-wave plate; HWP: half-wave plate; PZT: piezo-electric linear actuator; SNSPD: superconducting nanowire single-photon detector; TIA: time interval analyzer.}
\label{fig1}
\end{figure*}
\section{Theory and simulation}

We begin by presenting the theoretical model of the hybrid interferometer  as shown in Fig.\,\ref{fig1}. The two-photon state generated via spontaneous parametric down-conversion (SPDC) process can be described as:
\begin{small}
\begin{equation}\label{eq1}
\left| \psi \right\rangle = \int_0^{\infty}{\int_0^\infty {d\omega _s d\omega _i } } f(\omega _s ,\omega _i )\hat a_s^\dag (\omega _s )\hat a_i^\dag (\omega _i )\left| {00} \right\rangle,
\end{equation}
\end{small}
where $\omega$ is the angular frequency, $\hat a^\dag$ denotes the creation operator, and the subscripts $s$ and $i$ label the signal and idler photons, respectively. The function $f(\omega _s ,\omega _i )$ represents the joint spectral amplitude (JSA) of the signal and idler photons.

As derived in Appendix A, the coincidence probability $P(\tau_1,\tau_2)$ for the hybrid interferometer in Fig.\,\ref{fig1}(a) is given by:
\begin{small}
\begin{equation}\label{eq2}
\begin{split}
P(\tau_1,\tau_2)& =\int_{0}^{\infty}\int_{0}^{\infty}dt_3dt_4\left\langle\psi\left|\hat{E}_{3}^{(-)}\hat{E}_{4}^{(-)}\hat{E}_{4}^{(+)}\hat{E}_{3}^{(+)}\right|\psi\right\rangle\\&
=\frac1{16}\int_0^\infty\int_0^\infty d\omega_3d\omega_4 \Big|f(\omega_3,\omega_4)e^{-i\omega_3\tau_1} \\&
(e^{-i\omega_3\tau_2}+1)\left(e^{-i\omega_4\tau_2}+1\right)+f(\omega_4,\omega_3)e^{-i\omega_4\tau_1}
\\& (e^{-i\omega_4\tau_2}-1)(e^{-i\omega_3\tau_2}-1)\Big|^{2},
\end{split}
\end{equation}
\end{small}
where $\tau_1$ and $\tau_2$ are the optical path delays introduced before beam splitters BS1 and BS2, respectively.

We now introduce the frequency detunings $\Omega_{s,i}$ relative to half the pump frequency: $\omega_s$ =$\omega_p$/2 + $\Omega_s$ and $\omega_i$ = $\omega_p$/2 + $\Omega_i$ , where $\omega_p$ is the pump center frequency. Expressing the result in terms of the frequency sum $\Omega_+$ = $\Omega_s+\Omega_i$ and frequency difference $\Omega_-$ = $\Omega_s-\Omega_i$ , Eq.\,(\ref{eq2}) simplifies to
\begin{small}
\begin{equation}\label{eq4}
\begin{split}
P(\tau_1,\tau_2)& =\frac{1}{16}\iint d\Omega_{+}d\Omega_{-}\left|f_{+}(\Omega_{+})\right|^{2}\left|f_{-}(\Omega_{-})\right|^{2} \Big[8+\\&4\cos[(\omega_{p}+\Omega_{+})\tau_{2}]+4\cos(\Omega_{-}\tau_{2}) -2\operatorname{cos}[\Omega_{-}(\tau_{2}+\tau_{1})]\\&
-2\operatorname{cos}[\Omega_{-}(\tau_{2}-\tau_{1})] 
+2\cos[(\omega_p+\Omega_+)\tau_2+\Omega_-\tau_1]\\&+2\cos[(\omega_p+\Omega_+)\tau_2-\Omega_-\tau_1]\Big].
\end{split}
\end{equation}
\end{small}
In the above derivation, we have used the fact that the joint spectral amplitude can be treated as real and factorized in the sum- and difference-frequency coordinates. \rev{Specifically, the assumption of a real JSA $f(\omega_s,\omega_i) = f^*(\omega_s,\omega_i)$ is an acceptable convention. In our experiment, this convention is acceptable because the pump pulse is transform-limited with a flat spectral phase, and the type-II periodically
poled potassium titanyl phosphate (PPKTP) crystal is operated at phase-matching degeneracy with negligible dispersion effects within the pump bandwidth.} Under these conditions, the spectral phase of the biphoton is essentially constant, allowing the JSA to be treated as a real function. 

The assumption that $f(\Omega_+,\Omega_-) = f_+(\Omega_+) f_-(\Omega_-)$ is justified when the pump spectral envelope and the phase-matching function factorize in the sum- and difference-frequency coordinates. For our type-II SPDC source, this factorization is experimentally supported by the measured joint spectral intensity shown in Fig. 4(a), where is possible to observe that is aligned along the main diagonals, indicating that the JSA is separable along both the frequency-sum and frequency-difference directions.

Guided by these considerations, we take the joint spectral amplitude to be of the form $f(\Omega_{+},\Omega_{-})= f_+(\Omega_+) f_-(\Omega_-) =\exp(-\Omega_{+}^{2}/(2\sigma_{+}^{2}))\times \mathrm{sinc}(\Omega_{-}/\sigma_{-})$, where $\sigma_{\pm}$ denote the linewidth of two functions determined by the pump bandwidth and the phase-matching condition, respectively. The normalized coincidence
probability of Eq.\,(\ref{eq4}) now becomes
\begin{small}
\begin{equation}\label{eq5}
\begin{split}
P_{N}(\tau_1,\tau_2)=&1+\frac{1}{2}\cos(\omega_{p}\tau_{2})e^{-\tau_2^2\sigma_{+}^{2}}+\frac{1}{2}\Lambda(\sigma_{-}\tau_{2})\\
&-\frac{1}{4}\Lambda(\sigma_{-}(\tau_{2}+\tau_{1}))-\frac{1}{4}\Lambda(\sigma_{-}(\tau_{2}-\tau_{1})) \\&
+\frac{1}{2}\Lambda(\sigma_-\tau_1)\cos(\omega_p\tau_2)e^{-\tau_2^2\sigma_+^2},
\end{split}
\end{equation}
\end{small}
where $\Lambda$ is the triangular function, with zeros at $\pm 1/\sigma_{-}$. 
Here, the terms with $\cos(\omega_p\tau_2)$ constitute the N00N-state interference pattern, while the fourth and fifth terms correspond to HOM interference.

\begin{figure*}[!htp]
\centering
\includegraphics[width=1.0\textwidth]{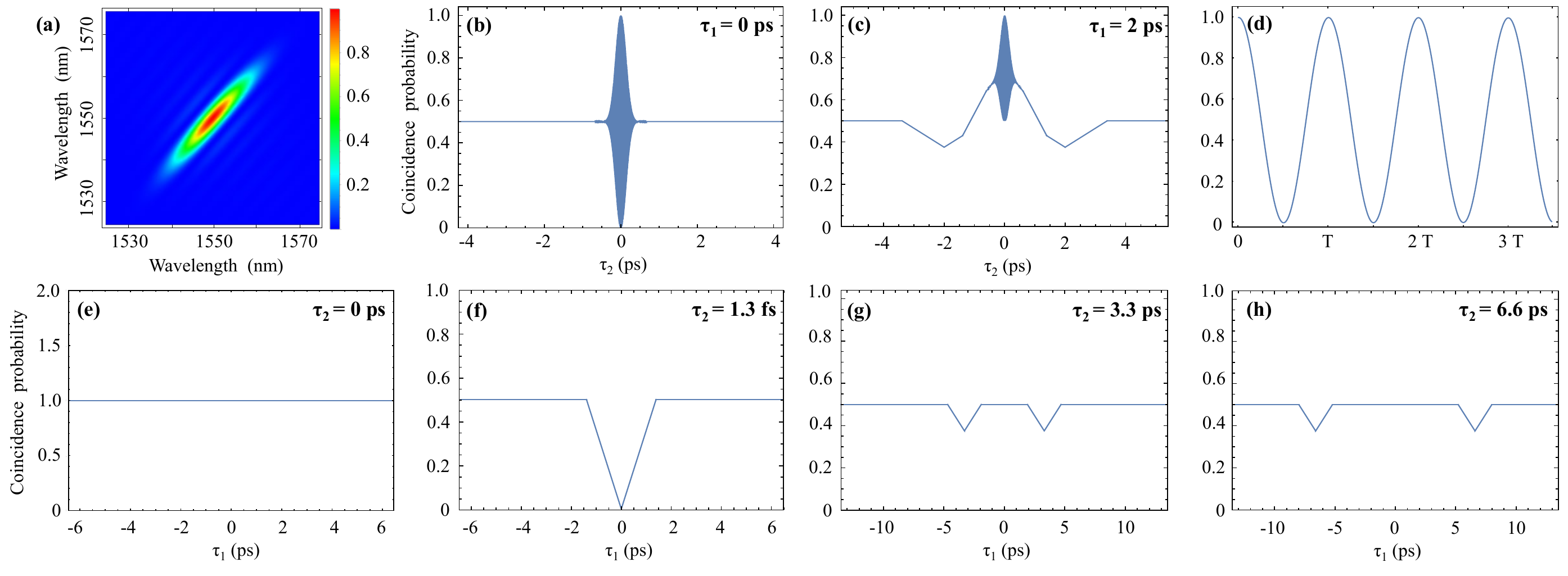}
\caption{Normalized theoretical simulation results. (a) Theoretical joint spectral intensity distribution for the type-II PPKTP crystal.
(b) Simulated coincidence probability versus $\tau_2$ with $\tau_1 = 0$~ps. 
(c) Simulated coincidence probability versus $\tau_2$ with $\tau_1 = 2$~ps.
(d) Magnified view around $\tau_2$ = 0 ps in figure (b), where T denotes the oscillation period.
(e-h) Simulated coincidence probability as a function of $\tau_1$ for fixed $\tau_2$: $\tau_2 = 0$~ps (e), $\tau_2 = 1.3$~fs (f), and $\tau_2 = 3.3$~ps (g), and $6.6$~ps (h). Here, panels (b), (f), and (c) correspond to pure N00N-state, HOM, and hybrid interference patterns, respectively.}
\label{fig2}
\end{figure*}

We then conducted systematic numerical simulations of the coincidence probability $P(\tau_1,\tau_2)$ as a function of the two time delays, based on a 10-mm-long type-II PPKTP crystal (with a poling period of 46.2 \textmu m) pumped by a Gaussian laser source centered at 775 nm. The results are presented in Fig.\,\ref{fig2}. Figure~\ref{fig2}(a) shows the theoretical joint spectral intensity (JSI) distribution for this crystal.
To elucidate the individual roles of the delay parameters, we first examine the system evolution with respect to $\tau_2$ for a fixed $\tau_1$. When $\tau_1$ = 0 ps, the signal and idler photons are perfectly indistinguishable. They undergo HOM interference at the first beam splitter (BS1 in Fig.\,\ref{fig1}(a)), resulting in photon bunching into a two-photon N00N-state $\ket{2002}$, which is then injected into the subsequent interferometer. Scanning 
$\tau_2$ under this condition yields a pure N00N-state interference pattern, as shown in Fig.\,\ref{fig2}(b). The oscillatory behavior is governed by the second term in Eq.\,\eqref{eq5} with a period of 
$T=\frac{2\pi}{\omega_{p}}$ $ \approx$ 2.6 fs. This period is half that expected for single-photon interference at the same wavelength, reflecting the phase sensitivity inherent to the two-photon N00N state. The fine structure of the fringes  around $\tau_2=0$ ps is further illustrated in Fig.\,\ref{fig2}(d).
When a non-zero delay is introduced (e.g., $\tau_1$ = 2 ps), the $\tau_2$-scan reveals a more complex pattern, shown in Fig.\,\ref{fig2}(c). While the oscillatory N00N-state background persists, two symmetric HOM dips emerge at $\tau_2=\mp\tau_1$. These dips originate from the combined effect of the fourth and fifth terms in Eq.\,\eqref{eq5}.
Physically, scanning $\tau_2$ can partially compensate for the path difference imposed by $\tau_1$, momentarily restoring photon indistinguishability and thereby inducing HOM interference at $\tau_2=\mp\tau_1$.
%

\begin{figure*}[!thp]
\centering
\includegraphics[width= 1.0\textwidth]{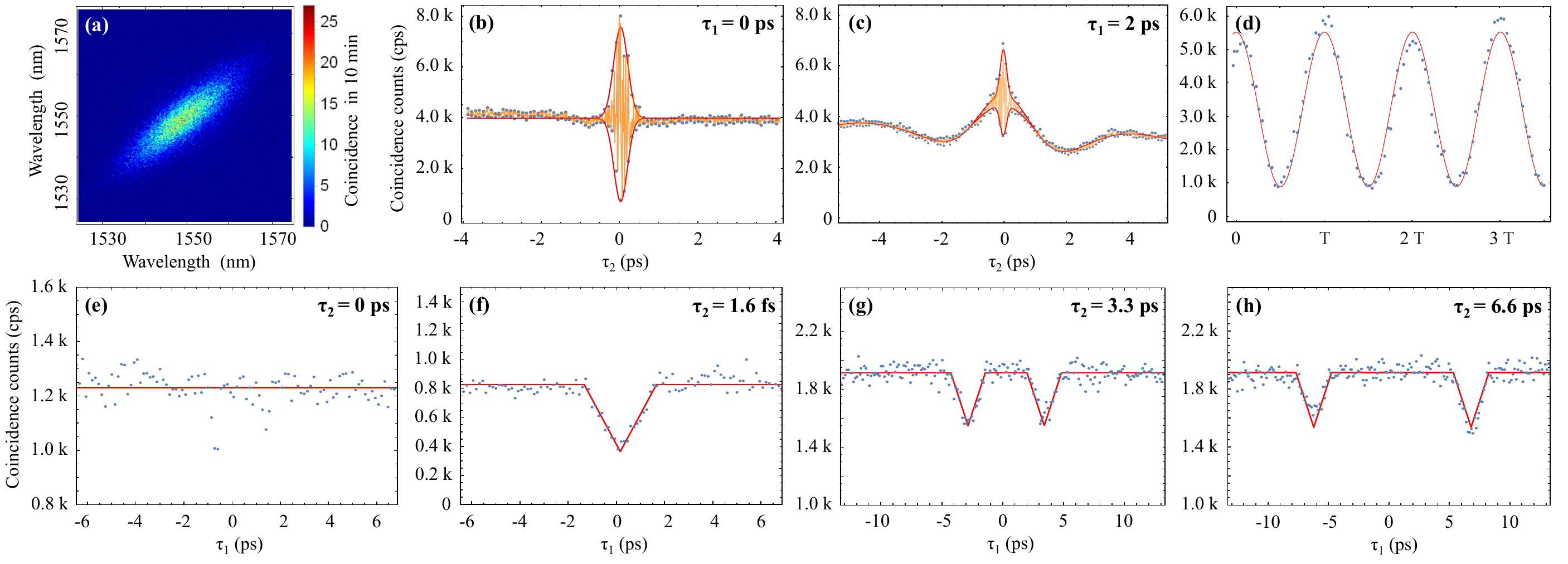}
\caption{Experimental results. (a) Measured joint spectral intensity of the type-II PPKTP crystal. (b–h) Coincidence counts as a function of optical delay. Blue dots represent experimental data; red solid curves denote theoretical fits. (b) Coincidence counts versus $\tau_2$ with $\tau_1$ = 0 ps. (c) Coincidence counts versus $\tau_2$ with $\tau_1$ = 2 ps. (d) Magnified view of (b) around $\tau_2$ = 0 ps, where T denotes the oscillation period. (e-h) Coincidence counts versus $\tau_1$ for fixed  $\tau_2$: $\tau_2$ = 0 fs (e), 1.6 fs (f), 3.3 ps (g), and 6.6 ps (h).}
\label{fig3}
\end{figure*}
%

Conversely, we analyze the system evolution by fixing $\tau_2$ and scanning $\tau_1$. When $\tau_2$ is set to a phase maximum of the N00N-state interference ($\tau_2$ = 0 fs), the system operates in a phase-sensitive regime. In this configuration, the HOM interference based on photon indistinguishability is fully suppressed, resulting in zero visibility, as shown in Fig.\,\ref{fig2}(e).
In contrast, setting $\tau_{2}$ to a phase minimum ($\tau_{2} = 1.3$ fs) maximizes the system's sensitivity to path information. Here, the delay $\tau_{1}$ directly controls the photons' temporal distinguishability, giving rise to a perfect HOM dip at $\tau_{1} = 0$ ps (Fig.\,\ref{fig2}(f)).
As $\tau_2$ increases further, the single HOM dip symmetrically splits into two, as depicted in Figure~\ref{fig2}(g-h). This splitting occurs because a larger $\tau_{2}$ enables the clear resolution of two independent quantum pathways within the detection time window. One pathway corresponds to both photons experiencing a total delay of $(\tau_{2} + \tau_{1})$, while the other corresponds to a total delay of $(\tau_{2} - \tau_{1})$. These two temporally distinguishable pathways each generate a HOM interference dip at $\tau_{1} = \mp \tau_{2}$, with a normalized depth of 0.25.

\section{Experiment and results}

The experimental setup for the hybrid interferometer is shown in Fig.\,\ref{fig1}(c). 
Femtosecond laser pulses with a repetition rate of 76 MHz, a bandwidth of 5.5 nm and a center wavelength of 775 nm were used to pump a 10-mm-long periodically poled KTiOPO$_4$ (PPKTP) crystal (poling period is 46.2 $\mu$m).
The PPKTP crystal was type-II phase matched (y$\to$y+z), and the signal and idler photons generated from SPDC process were orthogonally polarized \cite{Jin2022APL}.
After filtering by the long-pass filters and dichroic mirror, the biphotons were directed to two sequential time delay systems. The first system, which introduces the delay $\tau_1$, comprises a polarization beam splitter (PBS1) and a stepping motor. The second system, responsible for the delay $\tau_2$, consists of another polarization beam splitter (PBS2), a second stepping motor, and a piezo-electric (PZT) linear actuator.
Subsequently, the photons were directed to the third polarization beam splitter (PBS3). At PBS3, the orthogonally polarized photons were separated into two spatial paths according to their polarization states. Each output was then independently coupled into a single-mode optical fiber, which fed into a fiber spectrometer for joint spectral analysis.
The fiber spectrometer consisted of one cyclic 20-km-long single-mode fiber (SMF) for dispersion, two superconducting nanowire single-photon detectors (SNSPDs) \cite{Chi2018OL,Meng2022ACSPhotonics}, one synchronization signal from the laser, and a time interval analyzer (TIA). 

In this measurement scheme, the dispersive fiber maps the wavelength of each photon to its time of arrival at the SNSPD. By correlating these arrival times with the pump laser's synchronization signal via three-fold coincidence counting, the joint spectral intensity of the biphoton can be reconstructed. The experimentally measured JSI is shown in Fig.\,\ref{fig3}(a).

\begin{figure*}[!thp]
\centering
\includegraphics[width= 1.0\textwidth]{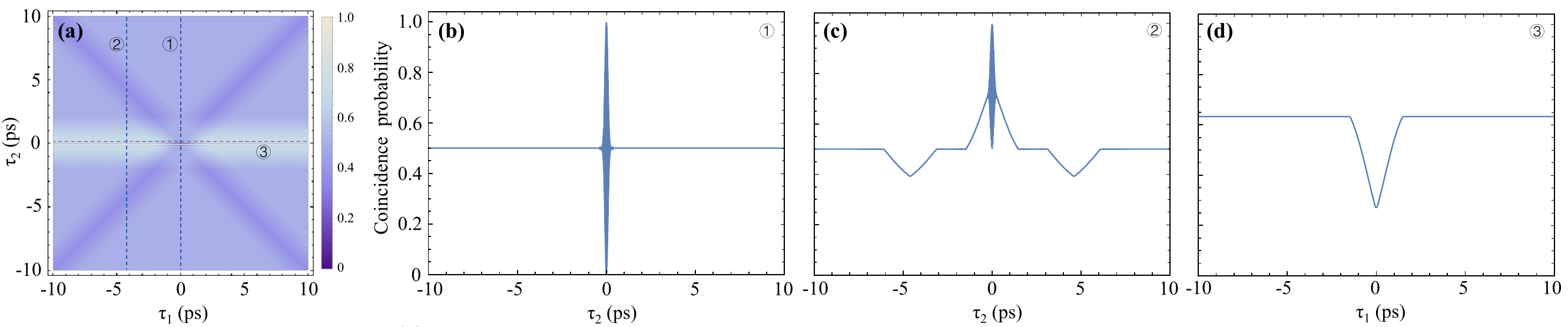}
\caption{(a) Density map of coincidence probability as a function of time delays, $\tau_1$ and $\tau_2$. (b-d) One-dimensional interference patterns corresponding to cross-sections 1, 2 and 3 in (a), respectively.}
\label{fig4}
\end{figure*}

Prior to the formal measurements, the reference positions corresponding to $\tau_1$ = 0 ps and $\tau_2$= 0 ps in the simulation were experimentally determined through a two-step calibration procedure. Firstly, the setup was configured for HOM interference by including only HWP1 in the optical path. Scanning the stepping motor controlling $\tau_1$ yielded a standard HOM dip, and its center was defined as the $\tau_1=0$ ps reference. Subsequently, $\tau_1$ is fixed at this position, HWP2 was inserted into the beam path to configure the system for N00N-state interference.  Scanning $\tau_2$ then produced a N00N-state interference fringe, and the position of its center was defined as the $\tau_2=0$ ps reference. During both calibration steps, all  half-wave plates were kept at fixed angles (e.g., 22.5°), making sure to complete the polarization transformation in the Fig.\,\ref{fig1}(b).
With the reference points established, the formal measurements were performed by setting $\tau_1$
to various values and scanning $\tau_2$.
The measured coincidence counts as a function of optical path delay are shown in Fig.\,\ref{fig3}(b) and (c), and the scan was performed using the stepping motor with a step size of 2 $\mu$m.
Fig.\,\ref{fig3}(b) exhibits the N00N-state interference pattern obtained with $\tau_1$ = 0 ps. When $\tau_1$ was increased to 2 ps by translating the corresponding stepper motor, the resulting pattern is shown in Fig.\,\ref{fig3}(c). To resolve the fine details of the N00N-state fringes in Fig.\,\ref{fig3}(b), a PZT with a step size of 40 nm was used for a high-resolution scan, yielding interference visibility of 75 $\pm$ 0.8$\%$ in Fig.\,\ref{fig3}(d).
The experimental results in Fig.\,\ref{fig3}(b-d) are consistent well with the corresponding simulations in Fig.\,\ref{fig2}(b-d). We next characterize the interferometer by fixing the delay $\tau_2$ at specific values and scanning the delay $\tau_1$. The measured coincidence counts as a function of optical path delay are presented in Fig.\,\ref{fig3}(e-h), the scanning of $\tau_1$ was performed using a stepping motor with a step size of 4 $\mu$m. For the data in Fig.\,\ref{fig3}(e) and (f), $\tau_2$ was set to 0 fs and 1.6 fs, respectively. These fine delays were implemented by precisely adjusting the PZT. This setting is obtained by moving the PZT by a small distance. In contrast, for the results shown in Fig.\,\ref{fig3}(g) and (h), $\tau_2$ was set to larger values of 3.3 ps and 6.6 ps, respectively, which were achieved by translating the second stepping motor. The experimental results are in excellent agreement with the corresponding simulation in Fig.\,\ref{fig2}(e-h). This agreement validates our theoretical model and confirms its predictive power for the system's behavior.

\section{Discussion}

To more intuitively illustrate the synergistic control of quantum interference by the two time delays $\tau_1$ and $\tau_2$, we present a two-dimensional density map of the coincidence probability as a function of both delays in Fig.\,\ref{fig4}(a). This map reveals a continuously tunable parameter space encompassing both phase-sensitive N00N-state interference fringes and path-indistinguishability-dominated HOM interference dips. 
By extracting one-dimensional interference curves along specific cross-sections (Fig.\,\ref{fig4}(b)-(d)), we can further elucidate the smooth transition between these two interference regimes. Firstly, the dotted line 1 in the density map ($\tau_1$ = 0) corresponds to pure N00N-state interference, shown in Fig.\,\ref{fig4}(b). Here, strong oscillatory fringes are centered at $\tau_2$ = 0 ps. Secondly, the dotted line 2 in the density map ($\tau_1$ = $-$4.6 ps), shown in Fig.\,\ref{fig4}(c), demonstrates a hybrid case. Here, the position of the N00N-state fringe pattern remains fixed, while its contrast diminishes and two distinct HOM dips emerge. These dips shift along the diagonal and anti-diagonal directions of the map as $\tau_1$ is varied. This indicates that the two interference mechanisms can be independently controlled and coexist within the same setup.
Finally, the dotted line 3 is obtained at a fixed $\tau_2$ = 50 fs, displaying a single HOM dip, shown in Fig.\,\ref{fig4}(f). Theoretically, the visibility of HOM interference is maximized at $\tau_2 = 1.3$ fs, as clearly evidenced in Fig.\,\ref{fig2}(f).
It is noteworthy that the experimentally observed optimal HOM dip (Fig.\,\ref{fig3}(f)) occurs at $\tau_2 = 1.6$ fs with a visibility of 54 $\pm$ 3.4$\%$. \rev{For reference, in the standard HOM configuration used to calibrate the $\tau_1 = 0$ ps reference (before inserting HWP2), we measured a raw visibility of approximately 91.7 $\pm$ 1.3$\%$ (See Appendix C for more details).} We have verified spectral degeneracy by temperature tuning and confirmed that accidental coincidences are negligible under our low pump power conditions. The remaining visibility reduction is mainly attributed to the difficulty of maintaining a stable lock of the PZT at the interference extremum due to environmental perturbations, which caused slow drift of the spatial mode overlap during the extended measurement time. This is a technical limitation rather than a fundamental one. We note that even with this non-optimal visibility, the key result---the dynamic tunability between HOM, N00N-state, and hybrid interference regimes within a single device---is unambiguously demonstrated and remains valid.

Overall, the dynamic tunability and coexistence of both interference mechanisms, visualized in Fig.\,\ref{fig4}, enable adaptive quantum sensing and establish a versatile platform for advanced quantum state characterization. More details about the sensitivity on the phase noise for HOM and N00N-state interferometry can be found in \cite{Yan2026arXiv}.

\section{Conclusion}
In summary, we theoretically and experimentally demonstrate a hybrid quantum interferometer that enables dynamic control between Hong-Ou-Mandel (HOM) and N00N-state interference by tuning the time delays  $\tau_1$ and $\tau_2$. The seamless transition between phase-insensitive, phase-sensitive, and simultaneous mixed interference is confirmed by excellent agreement between experimental results and theoretical simulations. This tunability allows a single device to adaptively meet the distinct demands of high-noise robustness and low-noise ultra-precise sensing. Our platform provides a versatile tool for advanced quantum state characterization and adaptive quantum metrology. This work paves the way for integrated quantum technologies capable of operating optimally across diverse and fluctuating environmental conditions.

\section*{Acknowledgments}
We thank Prof. Hui Jing and Prof. Ryosuke Shimizu for helpful discussions. This work is supported by the National Key R\&D Program (Grant No. 2024YFE0102400), the National Natural Science Foundation of China (Grants Nos. 12421005, 12574389,  92365106, 12074299, and 12074309), the Hunan Major Sci-Tech Program (Grant No. 2023ZJ1010), the Science and Technology Program of Hunan Province (Grant No. 2026QK3015), the Scientific Research Program Funded by Shaanxi Provincial Education Department (Program No. 24JR034), \rev{and the Natural Science Basic Research Plan in Shaanxi Province of China (Grant No. 2026JC-YBMS-0102).}
Dr. Dongzhou Wang was supported by  the Innovation Program for Quantum Science and Technology (2021ZD0300800). 
R.J.L.-M. thankfully acknowledges financial support by DGAPA-UNAM under
the project UNAM-PAPIIT IN101326.
Xiaolong Hu thankfully
acknowledges financial support from Steady Support
Fund for National Key Laboratory (JBS262800380).

\onecolumngrid
\clearpage
\renewcommand\thefigure{A\arabic{figure}}
\setcounter{figure}{0}
\setcounter{equation}{0}
\renewcommand\theequation{A\arabic{equation}}
\newpage
\section*{Appendix A: The coincidence probability $P(\tau_1,\tau_2)$ of the hybrid interferometer}
The two-photon state from a spontaneous parametric down-conversion (SPDC) process can be described as
\begin{small}
\begin{equation}\label{eqA1}
|\psi\rangle=\int_0^{\infty} d \omega_s \int_0^{\infty} d \omega_i f\left(\omega_s, \omega_i\right) \hat{a}_{s}^{\dagger}\left(\omega_s\right) \hat{a}_{i}^{\dagger}\left(\omega_i\right)|00\rangle,\\
\end{equation}
\end{small}
where $\omega$ is the angular frequency; $\hat a^\dag$ is the creation operator and the subscripts $s$ and $i$ denote the signal and idler photons from SPDC, respectively;$f(\omega _s ,\omega _i )$ is the joint spectral amplitude of the signal and idler photons.
The detection field operators of detector 1 (D3) and detector 2 (D4) are
\begin{small}
\begin{equation}\label{eqA2}
\hat E_3^{( + )} (t_3 ) = \frac{1}{{\sqrt {2\pi } }}\int_0^\infty  {d\omega _3 } \hat a_3 (\omega _3 )e^{ - i\omega _3 t_3 },
\end{equation}
\end{small}
\begin{small}
\begin{equation}\label{eqA3}
\hat E_4^{( + )} (t_4 ) = \frac{1}{{\sqrt {2\pi } }}\int_0^\infty  {d\omega _4 \hat a_4 (\omega _4 )} e^{ - i\omega _4 t_4 },
\end{equation}
\end{small}
where the subscripts $3$ and $4$ denote the photons detected by D3 and D4 respectively.
The transformation rule after the delay time $\tau_1$ and $\tau_2$ is
\begin{equation}\label{eqA4}
\begin{split}
\hat{a}_3\left(\omega_3\right)&=\frac{1}{\sqrt{2}}\left[\hat{a}_1(\omega_3)e^{-i\omega_3\tau_2}+\hat{a}_2(\omega_3)\right]\\&=\frac{1}{2}[\hat{a}_s(\omega_3)e^{-i\omega_3\tau_1}+\hat{a}_i(\omega_3)]e^{-i\omega_3\tau_2}\\&+[\hat{a}_s(\omega_3)e^{-i\omega_3\tau_1}-\hat{a}_i(\omega_3)],\\
\end{split}
\end{equation}


\begin{equation}
\begin{split}
\hat{a}_4(\omega_4)
&= \frac{1}{\sqrt{2}}\left[\hat{a}_1(\omega_4)e^{-i\omega_4\tau_2}-\hat{a}_2(\omega_4)\right] \\
&= \frac{1}{2}\bigl[\hat{a}_s(\omega_4)e^{-i\omega_4\tau_1}+\hat{a}_i(\omega_4)\bigr]e^{-i\omega_4\tau_2} \\
&\quad - \bigl[\hat{a}_s(\omega_4)e^{-i\omega_4\tau_1}-\hat{a}_i(\omega_4)\bigr].
\end{split}
\end{equation}

So, we can rewrite the field operators as
\begin{small}
\begin{equation}\label{eqA6}
\begin{split}
\hat{E}_{3}^{(+)}(t_{3})&=\frac{1}{\sqrt{2\pi}}\int_{0}^{\infty}d\omega_{3}\hat{a}_{3}(\omega_{3})e^{-i\omega_{3}t_{3}}
\\&=\frac{1}{2\sqrt{2\pi}}\int_{0}^{\infty}d\omega_{3}[\hat{a}_{s}(\omega_{3})(e^{-i\omega_{3}(\tau_{1}+\tau_{2})}\\&+e^{-i\omega_{3}\tau_{1}})e^{-i\omega_{3}t_{3}}+\hat{a}_{i}(\omega_{3})(e^{-i\omega_{3}\tau_{2}}-1)e^{-i\omega_{3}t_{3}}]
\end{split}
\end{equation}
\end{small}
and
\begin{small}
\begin{equation}\label{eqA7}
\begin{split}
\hat{E}_{4}^{(+)}(t_{4})&=\frac{1}{\sqrt{2\pi}}\int_{0}^{\infty}d\omega_{4}\hat{a}_{4}(\omega_{4})e^{-i\omega_{4}t_{4}}\\&
=\frac{1}{2\sqrt{2\pi}}\int_{0}^{\infty}d\omega_{4}[\hat{a}_{s}(\omega_{4})(e^{-i\omega_{4}(\tau_{1}+\tau_{2})}\\&-e^{-i\omega_{4}\tau_{1}})e^{-i\omega_{4}t_{4}}+\hat{a}_{i}(\omega_{4})(e^{-i\omega_{4}\tau_{2}}+1)e^{-i\omega_{4}t_{4}}].
\end{split}
\end{equation}
\end{small}

The coincidence probability P($\tau$), which is also the second-order correlation function $G_2$($\tau$), can be expressed as
\begin{small}
\begin{equation}\label{eqA8}
P(\tau_1,\tau_2) = \iint {d{t_3}d{t_4}} \left\langle {\psi \left| {\hat E_3^{( - )}\hat E_4^{( - )}\hat E_4^{( + )}\hat E_3^{( + )}} \right|\psi } \right\rangle. 
\end{equation}
\end{small}
Consider the $\hat E_4^{( + )}\hat E_3^{( + )}\left| \psi  \right\rangle$ , only 2 out of 4 terms exist.
The first term:
\begin{small}
\begin{equation}\label{eqA9}
\begin{aligned}
&\frac{1}{{8\pi }}\int_{0}^{\infty}  {\int_{0}^{\infty}  {d{\omega _3}d{\omega _4}} } {{\hat a}_s}({\omega _3})({e^{ - i{\omega _3}({\tau _1} + {\tau _2})}} + {e^{ - i{\omega _3}{\tau _1}}}){e^{ - i{\omega _3}{t_3}}} \\&\times {{\hat a}_i}({\omega _4})({e^{ - i{\omega _4}{\tau _2}}} + 1){e^{ - i{\omega _4}{t_4}}} \\& \times \int_{0}^{\infty}\int_{0}^{\infty}  {d{\omega _s}{d{\omega _i}} } f({\omega _s},{\omega _i})\hat a_s^\dag ({\omega _s})\hat a_i^\dag ({\omega _i})\left| {00} \right\rangle 
\\& = \frac{1}{{8\pi }}\int_{0}^{\infty}  {\int_{0}^{\infty}  {d{\omega _3}d{\omega _4}} } {e^{ - i{\omega _3}{t_3}}}{e^{ - i{\omega _4}{t_4}}}f({\omega _3},{\omega _4})\\&[{e^{ - i{\omega _3}{\tau _1}}}({e^{ - i{\omega _3}{\tau _2}}} + 1)({e^{ - i{\omega _4}{\tau _2}}} + 1)]\left| {00} \right\rangle. 
\end{aligned}
\end{equation}
\end{small}

In the above calculation, the equations of $\hat a(\omega )\hat a_{}^\dag ({\omega ^,})\left| 0 \right\rangle  = \delta (\omega  - {\omega ^,})\left| 0 \right\rangle $is used.
The second term:
\begin{small}
\begin{equation}\label{eqA10}
\begin{aligned}
& \frac{1}{{8\pi }}\int_{0}^{\infty}  {\int_{0}^{\infty}  {d{\omega _3}d{\omega _4}} } {{\hat a}_i}({\omega _3})({e^{ - i{\omega _3}{\tau _2}}} - 1){e^{ - i{\omega _3}{t_3}}} \\&\times {{\hat a}_s}({\omega _4})({e^{ - i{\omega _4}({\tau _1} + {\tau _2})}} - {e^{ - i{\omega _4}{\tau _1}}}){e^{ - i{\omega _4}{t_4}}} \\ & 
\times \int_{0}^{\infty}\int_{0}^{\infty}  {d{\omega _s}{d{\omega _i}} } f({\omega _s},{\omega _i})\hat a_s^\dag ({\omega _s})\hat a_i^\dag ({\omega _i})\left| {00} \right\rangle 
\\ & = \frac{1}{{8\pi }}\int_{0}^{\infty}  {\int_{0}^{\infty}  {d{\omega _3}d{\omega _4}} } {e^{ - i{\omega _3}{t_3}}}{e^{ - i{\omega _4}{t_4}}}f({\omega _4},{\omega _3})\\&[{e^{ - i{\omega _4}{\tau _1}}}({e^{ - i{\omega _4}{\tau _2}}} - 1)({e^{ - i{\omega _3}{\tau _2}}} - 1)]\left| {00} \right\rangle. 
\end{aligned}
\end{equation}
\end{small}
Combine these two terms: 
\begin{small}
\begin{equation}\label{eqA11}
\begin{split}
\hat E_4^{( + )}\hat E_3^{( + )}\left| \psi  \right\rangle  &= \frac{1}{{8\pi }}\int_{0}^{\infty}  {\int_{0}^{\infty}  {d{\omega _3}d{\omega _4}} } {e^{ - i{\omega _3}{t_3}}}{e^{ - i{\omega _4}{t_4}}}\\&[f({\omega _3},{\omega _4}){e^{ - i{\omega _3}{\tau _1}}}({e^{ - i{\omega _3}{\tau _2}}} + 1)({e^{ - i{\omega _4}{\tau _2}}} + 1) \\&
+ f({\omega _4},{\omega _3}){e^{ - i{\omega _4}{\tau _1}}}({e^{ - i{\omega _4}{\tau _2}}} - 1)({e^{ - i{\omega _3}{\tau _2}}} - 1)].
\end{split}
\end{equation}
\end{small}
Then,
{\footnotesize
\begin{equation}\label{eqA12}
\begin{aligned}
 &\left\langle {\psi \left| {\hat E_3^{( - )}\hat E_4^{( - )}\hat E_4^{( + )}\hat E_3^{( + )}} \right|\psi } \right\rangle  \\&=\frac{1}{{8\pi }}\int_{0}^{\infty}  {\int_{0}^{\infty}  {d{\omega _3}d{\omega _4}} } {e^{ - i{\omega _3}{t_3}}}{e^{ - i{\omega _4}{t_4}}}\\&[f({\omega _3},{\omega _4}){e^{ - i{\omega _3}{\tau _1}}}({e^{ - i{\omega _3}{\tau _2}}} + 1)({e^{ - i{\omega _4}{\tau _2}}} + 1) \\&
+ f({\omega _4},{\omega _3}){e^{ - i{\omega _4}{\tau _1}}}({e^{ - i{\omega _4}{\tau _2}}} - 1)({e^{ - i{\omega _3}{\tau _2}}} - 1)]
\\ & \times \frac{1}{{8\pi }}\int_{0}^{\infty}  {\int_{0}^{\infty}  {d{\omega^,}_3d{\omega ^,}_4} } {e^{i{\omega ^,}_3{t_3}}}{e^{i{\omega ^,}_4{t_4}}}\\&[f({\omega ^,}_3,{\omega ^,}_4){e^{i{\omega _3}{\tau _1}}}({e^{i{\omega ^,}_3{\tau _2}}} + 1)({e^{i{\omega ^,}_4{\tau _2}}} + 1) \\&
+ f({\omega ^,}_4,{\omega ^,}_3){e^{i{\omega ^,}_4{\tau _1}}}({e^{i{\omega ^,}_4{\tau _2}}} - 1)({e^{i{\omega ^,}_3{\tau _2}}} - 1)].
\end{aligned}
\end{equation}
}
Finally
{\footnotesize
\begin{equation}\label{eqA13}
\begin{split}
P(\tau_1,\tau_2) &= \iint {d{t_3}d{t_4}} \left\langle {\psi \left| {\hat E_3^{( - )}\hat E_4^{( - )}\hat E_4^{( + )}\hat E_3^{( + )}} \right|\psi } \right\rangle  \\&
= \frac{1}{{16}}\iint  d{\omega _3}d{\omega _3^,}\delta ({\omega _3} - \omega _3^,)\iint  {d{\omega _4}} d\omega _4^,\delta ({\omega _4} - \omega _4^,)\\&
\times [f({\omega _3},{\omega _4}){e^{ - i{\omega _3}{\tau _1}}}({e^{ - i{\omega _3}{\tau _2}}} + 1)({e^{ - i{\omega _4}{\tau _2}}} + 1) \\&+ f({\omega _4},{\omega _3}){e^{ - i{\omega _4}{\tau _1}}}({e^{ - i{\omega _4}{\tau _2}}} - 1)({e^{ - i{\omega _3}{\tau _2}}} - 1)]\\&
\times [f({\omega ^,}_3,{\omega ^,}_4){e^{i{\omega _3}{\tau _1}}}({e^{i{\omega ^,}_3{\tau _2}}} + 1)({e^{i{\omega ^,}_4{\tau _2}}} + 1) \\&+ f({\omega ^,}_4,{\omega ^,}_3){e^{i{\omega ^,}_4{\tau _1}}}({e^{i{\omega ^,}_4{\tau _2}}} - 1)({e^{i{\omega ^,}_3{\tau _2}}} - 1)]\\&
= \frac{1}{{16}}\iint  {d{\omega _3}}  d{\omega _4}\left| [f({\omega _3},{\omega _4}){e^{ - i{\omega _3}{\tau _1}}}({e^{ - i{\omega _3}{\tau _2}}} + 1)\right.\\&\left.({e^{ - i{\omega _4}{\tau _2}}} + 1)+ f({\omega _4},{\omega _3}){e^{ - i{\omega _4}{\tau _1}}}\right.\\&\left.({e^{ - i{\omega _4}{\tau _2}}} - 1)({e^{ - i{\omega _3}{\tau _2}}} - 1)] \right|^2.
\end{split}
\end{equation}}

For simplification, we change $\omega _3$ to $\omega _s$; and $\omega _4$ to  $\omega _i$
\begin{small}
\begin{equation}\label{eqA14}
\begin{split}
P(\tau_1,\tau_2)  &= \frac{1}{{16}}\iint {d{\omega _s}}  d{\omega _i}\left| [f({\omega _s},{\omega _i}){e^{ - i{\omega _s}{\tau _1}}}({e^{ - i{\omega _s}{\tau _2}}} + 1)\right.\\&\left.({e^{ - i{\omega _i}{\tau _2}}} + 1) + f({\omega _i},{\omega _s}){e^{ - i{\omega _i}{\tau _1}}}({e^{ - i{\omega _i}{\tau _2}}} - 1)\right.\\&\left.({e^{ - i{\omega _s}{\tau _2}}} - 1)] \right|^2.
\end{split}
\end{equation}
\end{small}

We assume that f=f* and f is a real number, then rewrite Eq.\,(\ref{eqA14}) to get the following formula:
\begin{small}
\begin{equation}\label{eqA15}
\begin{aligned}
P(\tau_1,\tau_2)& =\iint dt_{3}dt_{4}\left\langle\psi\left|\hat{E}_{3}^{(-)}\hat{E}_{4}^{(-)}\hat{E}_{4}^{(+)}\hat{E}_{3}^{(+)}\right|\psi\right\rangle  \\
&=\frac{1}{16}\int_{0}^{\infty}\int_{0}^{\infty}d\omega_{3}d\omega_{4}\left|f(\omega_{3},\omega_{4})\right|^{2}\\&[8+4\cos[(\omega_{3}+\omega_{4})\tau_{2}]\\&+4\cos[(\omega_{3}-\omega_{4})\tau_{2}]-2\cos[(\omega_3-\omega_4)(\tau_2+\tau_1)]\\
&-2\cos[(\omega_3-\omega_4)(\tau_2-\tau_1)]\\&
+2\cos[(\omega_3+\omega_4)\tau_2+(\omega_3-\omega_4)\tau_1]\\
&+2\cos[(\omega_3+\omega_4)\tau_2-(\omega_3-\omega_4)\tau_1]].
\end{aligned}
\end{equation}
\end{small}

Assuming that $\omega_s$ =$\omega_p$/2 + $\Omega_s$ and $\omega_i$ = $\omega_p$/2 + $\Omega_i$, where $\Omega_{s,i}$ is the frequency detuning between the signal (idler)photon and the half of pump center frequency $\omega_p$/2, then Eq.\,(\ref{eqA15}) can be expressed as the  form of frequency sum $\Omega_+$ = $\Omega_s$ + $\Omega_i$ and frequency difference $\Omega_-$ = $\Omega_s$ - $\Omega_i$.
\begin{small}
\begin{equation}\label{eqA16}
\begin{aligned}
P(\tau_1,\tau_2)& =\frac{1}{16} \int_{0}^{\infty} \int_{0}^{\infty}d\Omega_{+}d\Omega_{-} | f_{+}(\Omega_{+}) |^{2} | f_{-}(\Omega_{-}) |^{2} \\
&[8+4 \cos[(\omega_{p}+\Omega_{+})\tau_{2}]+4 \cos[\Omega_{-}\tau_{2}] \\
&-2\cos[\Omega_-(\tau_2+\tau_1)]-2\cos[\Omega_-(\tau_2-\tau_1)] \\
&+2\cos[(\omega_p+\Omega_+)\tau_2+\Omega_-\tau_1]\\
&+2\cos[(\omega_p+\Omega_+)\tau_2-\Omega_-\tau_1]],
\end{aligned}
\end{equation}
\end{small}

which corresponds to Eq.\,(3) in the main text. In general, the joint spectral amplitude cannot be factorized as a product of separate functions of $\omega_s$ and $\omega_i$. However, when expressed in terms of the collective coordinates $\Omega_+$ and $\Omega_-$, the JSA can be factorized as $f(\Omega_+,\Omega_-) = f_+(\Omega_+) f_-(\Omega_-)$, since the pump envelope function depends only on $\Omega_+$, while the phase-matching function depends only on $\Omega_-$ \cite{Jin2018PRAppl,Li2025PRL}. This factorization allows the integral in Eq.\,(\ref{eqA16}) to be evaluated independently with respect to $\Omega_+$ and $\Omega_-$, leading to the normalized coincidence probability. Next, we take JSA as $f(\Omega_{+},\Omega_{-})=\exp(-\frac{\Omega_{+}^{2}}{2\sigma_{+}^{2}})
\mathrm{sinc}\frac{\Omega_{-}}{\sigma_{-}}$, where $\sigma_{\pm}$ denote the linewidth of two functions determined by the linewidth of  pump pulse and the phase-matching condition, respectively. Eq.\,(\ref{eqA16}) now becomes

\begin{small}
\begin{equation}\label{eqA17}
\begin{aligned}
P(\tau_1,\tau_2) & =\frac{1}{16}[8\times2\pi\times\sqrt{\pi}\sigma_{+}^{}\sigma_{-}^{}+4\times2\pi\times\sqrt{\pi}\sigma_{+}^{}\sigma_{-}^{}\\  & \cos(\omega_{p}\tau_2)e^{-\tau_2^2\sigma_{+}^2}+4\times2\pi\times\sqrt{\pi}\sigma_{+}^{}\sigma_{-}^{}\Lambda(\sigma_{-}^{}\tau_2)\\  & -2\times2\pi\times\sqrt{\pi}\sigma_{+}^{}\sigma_{-}^{}\Lambda(\sigma_{-}^{}(\tau_2+\tau_1))\\  & -2\times2\pi\times\sqrt{\pi}\sigma_{+}^{}\sigma_{-}^{}\Lambda(\sigma_{-}^{}(\tau_2-\tau_1))\\  & +2\times2\pi\times\sqrt{\pi}\sigma_{+}^{}\sigma_{-}^{}\cos(\omega_{p}\tau_2)e^{-\tau_2^2\sigma_{+}^2}\Lambda(\sigma_{-}^{}\tau_1)\\  & +2\times2\pi\times\sqrt{\pi}\sigma_{+}^{}\sigma_{-}^{}\cos(\omega_{p}\tau_2)e^{-\tau_2^2\sigma_{+}^2}\Lambda(\sigma_{-}^{}\tau_1)].
\end{aligned}
\end{equation}
\end{small}

In the above equation, $\Lambda$ is the triangular function with zeros at $\pm 1/\sigma_{-}$, $2\pi\times\sqrt{\pi}\sigma_{+}^{}\sigma_{-}^{}$ is the result of the integral of $f(\omega_{3},\omega_{4})$ over time. Here, we treat this integral result as a normalized coefficient, then Eq.\,(\ref{eqA17}) can be normalized as
\begin{small}
\begin{equation}\label{eqA18}
\begin{aligned}
P_{N}(\tau_1,\tau_2)&=1+\frac{1}{2}\cos(\omega_{p}\tau_{2})e^{-\tau_2^2\sigma_{+}^{2}}+\frac{1}{2}\Lambda(\sigma_{-}\tau_{2})\\  & -\frac{1}{4}\Lambda(\sigma_{-}(\tau_{2}+\tau_{1}))-\frac{1}{4}\Lambda(\sigma_{-}(\tau_{2}-\tau_{1})) \\&
+\frac{1}{2}\Lambda(\sigma_-\tau_1)\cos(\omega_p\tau_2)e^{-\tau_2^2\sigma_+^2}.
\end{aligned}
\end{equation}
\end{small}

In the Eq.\,(\ref{eqA18}), the second term corresponds to N00N-state interference with an oscillation period $\frac{2\pi}{\omega_{p}}$. The third term corresponds to a HOM peak at the center of $\tau_2=0$, as shown in Fig.\,\ref{SI-Fig1}(b). And the 
The fourth and fifth terms correspond to the HOM dip of $\tau_2=\mp\tau_1$, respectively. The width of the above HOM interference is inversely proportional to the $\sigma_{-}$.

\section*{Appendix B: The formation of interference pattern}

\begin{figure*}[!thp]
\centering
\includegraphics[width= 0.95\textwidth]{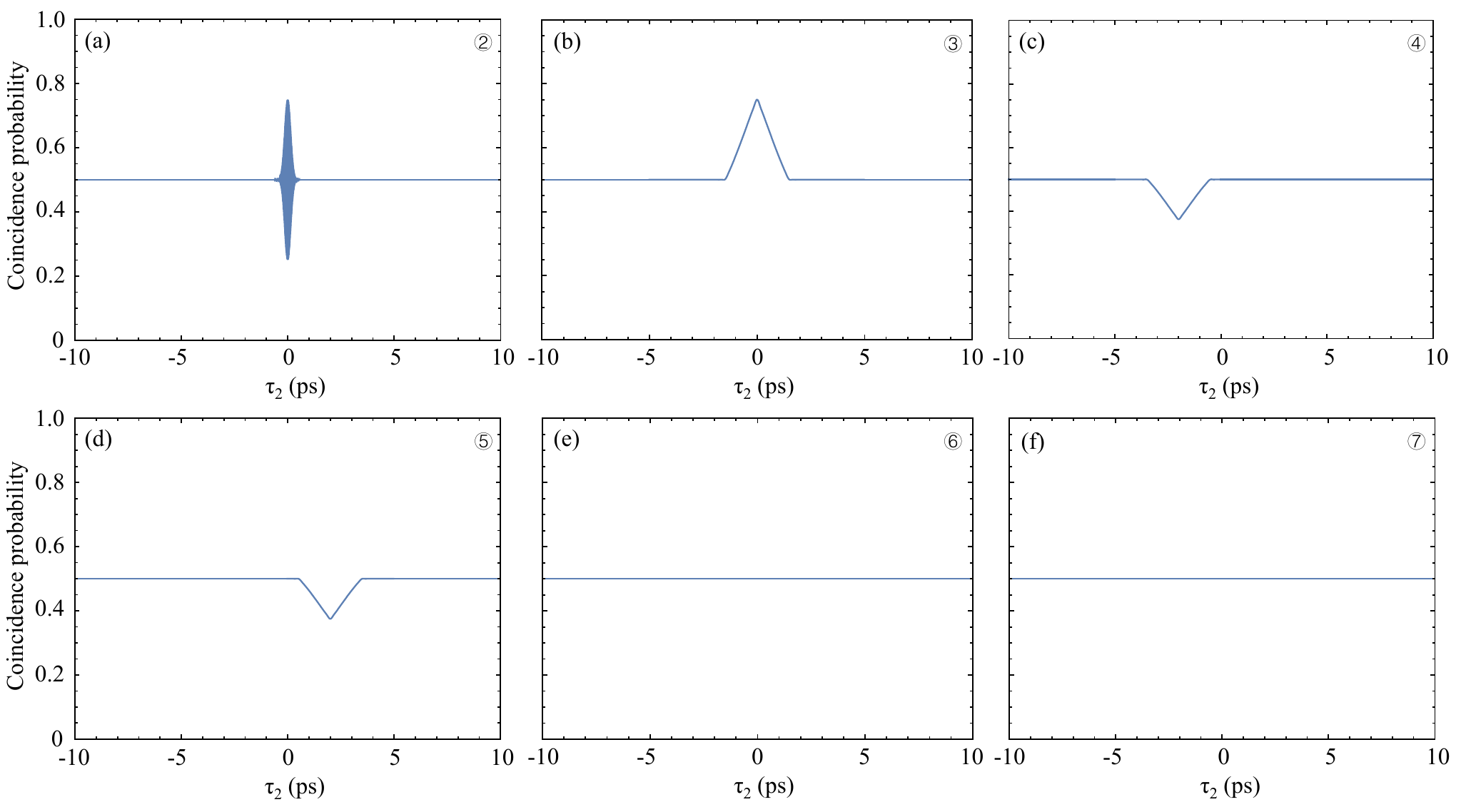}
\caption{(a-f) The interference pattern from the second term to the seventh term of $P(\tau_1,\tau_2)$ in the Eq.\,(3), respectively.}
\label{SI-Fig1}
\end{figure*}

Here, we will study the formation of interference fringe shapes. In the Eq.\,(\ref{eq4}), P($\tau$) has seven terms. The first term of the integral is the normalization factor. In order to fully understand this formula, we take $\tau_1$= 2 ps as an example, and draw the images of the second term to the seventh term respectively, corresponding to Fig.\,\ref{SI-Fig1}(a-f). As we can see, the second term is the sum frequency term, which produces the interference of N00N-states related to $\tau_2$. The third term is the difference frequency term, which produces bunching peak related to $\tau_2$. These two terms are only related to $\tau_2$. Therefore, when $\tau_1$ changes, it has no influence on these two terms, and the N00N-state interference fringe and the peak corresponding to these two terms will not move. The fourth and fifth terms are difference frequency terms, which are used to generate HOM dip, and are influenced by $\tau_1$ and $\tau_2$. When $\tau_1$ changes, dip will move left and right. The sixth and seventh integrals are always 0.5. It is not difficult to see that all the superimposed images are consistent with the theoretical Fig.\,\ref{fig2}(c), which is helpful for us to fully understand the corresponding changes of HOM interference and N00N-state interference under the action of two time delays.

\rev{\section*{Appendix C: Standard HOM Interference in the Calibration}

During the calibration procedure described in the main text, we performed a standard HOM interference measurement to establish the reference point at $\tau_1 = 0$~ps. In accordance with the setup shown in Fig.~\ref{fig1}, HWP2 was removed from the optical path while HWP1 remained in place, and the PZT was held stationary to define a fixed reference position. The resulting HOM dip is presented in Fig.~\ref{HOM}.

\begin{figure}[!h]
\centering
\includegraphics[width=0.45\textwidth]{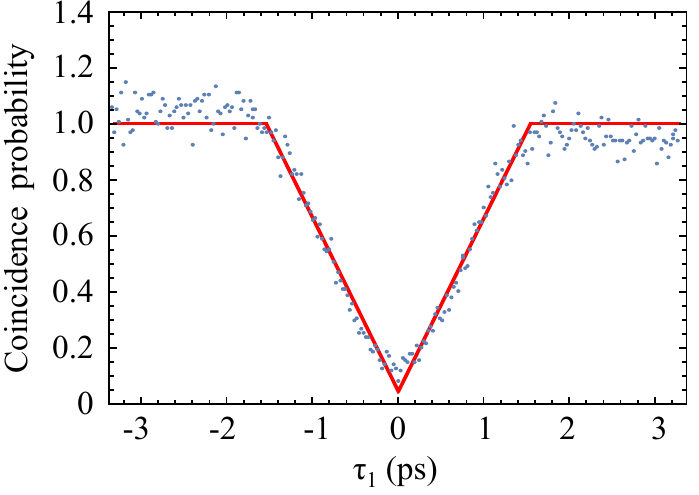}
\rev{\caption{Normalized  HOM interference dip measured during the calibration procedure (with HWP2 removed and HWP1 in place). Blue dots are experimental data points and the red solid curve is the theoretical fit. }
\label{HOM}}
\end{figure}

Without subtracting accidental coincidence counts, the raw visibility is $91.7 \pm 1.3\%$. After fitting with a triangular function, the fitted visibility improves to $95.5\%$. This high interference visibility attests to the excellent indistinguishability of our photon source, as well as the precise spatial and temporal overlap achieved during the calibration.

}


\end{document}